\documentclass[final,5p,times,twocolumn]{elsarticle}
\journal{Phys. Lett. B}

\usepackage{graphicx,color,rotating,pifont}
\usepackage{amsmath,amssymb,bm}

\DeclareMathSymbol{\NS}{\mathord}{AMSb}{"4E}

\newcommand{\ket}[1]{\ensuremath{\,|{#1}\rangle}}
\newcommand{\braket}[2]{\ensuremath{\langle{#1}|{#2}\rangle}}
\newcommand{\matrixe}[3]{\ensuremath{\langle{#1}|\,{#2}\,|{#3}\rangle}}

\newcommand{\dmatrixe}[2]{\matrixe{#1}{#2}{#1}}

\newcommand{\expect}[1]{\ensuremath{\langle{#1}\rangle}}

\newcommand{\comm}[2]{\ensuremath{[{#1},{#2}]}}

\newcommand{\op}[1]{\ensuremath{#1}}
\newcommand{\adj}[1]{\ensuremath{{{#1}}^{\dag}}}

\renewcommand{\vec}[1]{\ensuremath{\bm{#1}}}

\newcommand{\partd}[2]{\ensuremath{ \frac{\partial {#1}}{\partial {#2}} }}

\newcommand{\cO}{\ensuremath{\op{c}}}

\newcommand{\tO}{\ensuremath{\op{t}}}
\newcommand{\vO}{\ensuremath{\op{v}}}

\newcommand{\ccO}{\ensuremath{\adj{\op{c}}}}

\newcommand{\AO}{\ensuremath{\hat{\op{A}}}}

\newcommand{\HO}{\ensuremath{\op{H}}}

\newcommand{\PO}{\ensuremath{\op{P}}}
\newcommand{\TO}{\ensuremath{\op{T}}}

\newcommand{\VO}{\ensuremath{\op{V}}}

\newcommand{\pOV}{\ensuremath{\vec{\op{p}}}}
\newcommand{\qOV}{\ensuremath{\vec{\op{q}}}}

\newcommand{\Tint}{\ensuremath{\TO_\text{int}}}

\newcommand{\Hint}{\ensuremath{\HO_\text{int}}}

\newcommand{\aHO}{\ensuremath{a_{\text{HO}}}}

\newcommand{\nuc}[2]{\ensuremath{^{#2}\mathrm{#1}}}

\newcommand{\keV}{\ensuremath{\,\text{keV}}}

\newcommand{\eV}{\ensuremath{\,\text{eV}}}

\newcommand{\tr}{\ensuremath{\mathrm{tr}}}

\newcommand{\symboldiamond}[1][black]{({\color{#1}$\blacklozenge$})}

\newcommand{\symbolbox}[1][black]{({\color{#1}$\blacksquare$})}
\newcommand{\symbolcircle}[1][black]{({\color{#1}\ding{108}})}

\newcommand{\symbolcross}[1][black]{({\color{#1}\ding{58}})}

\definecolor{FGViolet}{rgb}{0.61,0.32,0.61}
\definecolor{FGDarkBlue}{rgb}{0,0,0.6}
\definecolor{FGBlue}{rgb}{0,0,0.8}
\definecolor{FGLightBlue}{rgb}{0.2, 0.6, 0.8}
\definecolor{FGGreen}{rgb}{0.2,0.7,0.2}
\definecolor{FGLightGreen}{rgb}{0.4,1,0.4}
\definecolor{FGYellow}{rgb}{1,0.95,0}
\definecolor{FGOrange}{rgb}{0.95,0.5,0.1}
\definecolor{FGRed}{rgb}{0.8,0,0}
\definecolor{FGWhite}{rgb}{1,1,1}
\definecolor{FGLightGray}{rgb}{0.8,0.8,0.8}
\definecolor{FGGray}{rgb}{0.5,0.5,0.5}
\definecolor{FGDarkGray}{rgb}{0.3,0.3,0.3}
\definecolor{FGBlack}{rgb}{0,0,0}

\begin{document}
\begin{frontmatter}
\title{Treatment of the Intrinsic Hamiltonian in Particle-Number Nonconserving Theories}

\author[ikp]{H. Hergert}
\ead{Heiko.Hergert@physik.tu-darmstadt.de}

\author[ikp]{R. Roth}
\ead{Robert.Roth@physik.tu-darmstadt.de}

\address[ikp]{Institut f\"ur Kernphysik, Technische Universit\"at Darmstadt, Schlossgartenstr. 9, 
64289 Darmstadt, Germany}

\date{\today}

\begin{abstract}
We discuss the implications of using an intrinsic Hamiltonian in theories without particle-number conservation, e.g., the Hartree-Fock-Bogoliubov approximation, where the Hamiltonian's particle-number dependence leads to discrepancies if one naively replaces the particle-number operator by its expectation value. We develop a systematic expansion that fixes this problem and leads to an a posteriori justification of the widely-used one- plus two-body form of the intrinsic kinetic energy in nuclear self-consistent field methods. The expansion's convergence properties as well as its practical applications are discussed for several sample nuclei.
\end{abstract}

\begin{keyword}
  nuclear structure \sep intrinsic Hamiltonian \sep Hartree-Fock-Bogoliubov \sep particle-number conservation
  \PACS 21.60.Jz 
\end{keyword}

\end{frontmatter}

\section{Introduction}
Symmetry breaking is a powerful concept in quantum many-body theory. In nuclear phyiscs, well-known examples are the breaking of the rotational and translational symmetries of the many-body Hamiltonian by the use of localized single-particle states in the construction of the many-body Hilbert space; the latter, in particular, can cause sizable center-of-mass contaminations of the energies and the many-body wave functions unless the symmetry is restored (see e.g. \cite{Ring:1980,Dobaczewski:2009ie} and Refs. therein). In contrast, the breaking of the particle- number symmetry by quasi-particle methods like the Hartree-Fock-Bogoliubov (HFB) \cite{Ring:1980} approach is a useful tool because it leads to an efficient treatment of the important nuclear pairing correlations, although one will ultimately want to restore this symmetry in a finite system like the nucleus.

Since nuclei are self-bound objects the proper starting point of a nuclear many-body calculation is the translationally invariant intrinsic Hamiltonian
\begin{equation}
   \Hint = \TO-\TO_\text{cm} + \VO = \Tint + \VO\,.
\end{equation}
The use of the intrinsic kinetic energy $\Tint$ in a simple Hartree-Fock (HF) calculation has consequences for the validity of Koopmans' theorem, and thereby the interpretation of the HF eigenvalues as single-particle energies. A detailed analysis was given by Khadkikar and Kamble in Ref. \cite{Khadkikar:1974} and referenced repeatedly over the past few decades (see e.g. \cite{Jaqua:1992} or our own work \cite{Roth:2005ah}). However, this analysis makes explicit use of the properties of the HF Slater determinant $\ket{\Psi}$, including the assumption that it is an eigenstate of the particle-number operator with eigenvalue $A$:
\begin{equation}\label{eq:A_eigen}
   \AO\ket{\Psi}=A\ket{\Psi}\,.
\end{equation}
Naturally, this condition does not hold in a method like HFB, where the particle number is not conserved. We will analyze this case in detail in the following.

\section{Theory}
\subsection{The General Case}
Since we want to deal with theories that do not conserve particle number, we consider operators in \emph{Fock} space. In this case, the intrinsic kinetic energy operator can be expressed either as a sum of one- and two-body operators,
\begin{equation}
   \Tint^{(a)}=\left(1-\frac{1}{\AO}\right)\sum_i\frac{\pOV_i^2}{2m}-\frac{1}{\AO m}\sum_{i<j}\pOV_i\cdot\pOV_j\label{eq:tint_obtb}
\end{equation}
or a sum of two-body operators alone, i.e., the relative kinetic energies of each nucleon pair:
\begin{equation}
  \Tint^{(b)}=\frac{2}{\AO}\sum_{i<j}\frac{\qOV_{ij}^2}{2\mu}=\frac{1}{2\AO}\sum_{i<j}\frac{(\pOV_i-\pOV_j)^2}{m}\label{eq:tint_rel}\,.
\end{equation}
The equality of these two expressions follows from the relation
\begin{align}
   \sum_{i<j}\left(\pOV_i-\pOV_j\right)^2&=\sum_{i<j}\left(\pOV_i^2+\pOV_j^2-2\pOV_i\cdot\pOV_j\right)\notag\\
	&=\left(\AO-1\right)\sum_i\pOV_i^2-2\sum_{i<j}\pOV_i\cdot\pOV_j\label{eq:A_promotion}\,,
\end{align}
where the $\AO$ resulting from the summation over the second independent particle $i$ or $j$ in the first two terms is again a \emph{Fock space operator} measuring the total particle number. This distinction is inconsequential as long as one works in a Hilbert space with \emph{fixed particle number}, because $\AO$ can then be replaced by the corresponding eigenvalue. Naturally, one is tempted to use a similar replacement $\AO\to\expect{\AO}$ in Fock space as well, but we will demonstrate in the following that this naive treatment of the particle-number dependence leads to discrepancies.

Consider a many-body state $\ket{\Psi}$ without fixed particle number. Taking the energy expectation value of the intrinsic Hamiltonian with Eqs. \eqref{eq:tint_obtb} or \eqref{eq:tint_rel} in this state, we obtain the energy functionals
\begin{equation}\label{eq:E_obtb}
   E^{(a)}[\Psi]= \expect{\left(1-\frac{1}{\AO}\right)\TO}-\frac{1}{m}\sum_{i<j}\expect{\frac{1}{\AO}\pOV_i\cdot\pOV_j} + \expect{V}
\end{equation}
and 
\begin{equation}\label{eq:E_rel}
   E^{(b)}[\Psi]= \frac{1}{2m}\sum_{i<j}\expect{\frac{1}{\AO}\left(\pOV_i-\pOV_j\right)^2} + \expect{V}\,,
\end{equation}
respectively. Since $\ket{\Psi}$ does not satisfy the eigenvalue equation \eqref{eq:A_eigen}, we have to consider the operator $\AO^{-1}$ directly in all expectation values, which will require a series expansion in practice. To this end, we note that
\begin{equation}\label{eq:A_series}
   \frac{1}{\AO}=\frac{1}{\expect{\AO}}\frac{1}{1+\frac{\Delta\AO}{\expect{\AO}}}=\frac{1}{\expect{\AO}}\sum_n(-1)^n\left(\frac{\Delta\AO}{\expect{\AO}}\right)^n\,,
\end{equation}
where we have introduced
\begin{equation}\label{eq:DeltaA}
   \Delta\AO\equiv\AO-\expect{\AO}\,.
\end{equation}
Eq. \eqref{eq:A_series} defines a formal expansion of the energy functionals in powers of $\expect{\AO}^{-1}$. 
Applying this expansion to $E^{(a)}$, we obtain
\begin{align}
   E^{(a)}&=\left(1-\frac{1}{\expect{\AO}}\right)\expect{\TO}-\frac{1}{\expect{\AO}m}\sum_{i<j}\expect{\pOV_i\cdot\pOV_j}+\expect{V}\notag\\[3pt]
      &\hphantom{=}+\frac{\expect{\TO\Delta\AO}}{\expect{\AO}^2}+\frac{1}{\expect{\AO}^2m}\sum_{i<j}\expect{\pOV_i\cdot\pOV_j\Delta\AO}+O\left(\expect{\AO}^{-3}\right)\,.\label{eq:E_obtb_expansion}
\end{align}
Denoting truncations of this series containing all terms up to a given order $k$ by $E^{(a)}_k$, the leading (LO) and next-to-leading order (NLO) functionals are
\begin{align}
   E^{(a)}_0&=\expect{\TO} + \expect{\VO}\,,\\
   E^{(a)}_1&=E^{(a)}_0-\frac{1}{\expect{\AO}}\expect{\TO}-\sum_{i<j}\frac{\expect{\pOV_i\cdot\pOV_j}}{\expect{\AO}m}\,.\label{eq:E1_nlo}
\end{align}
We note that the NLO functional is the one we would obtain by naively replacing $\AO$ with $\expect{\AO}$ in Eq. \eqref{eq:E_obtb}.

Plugging the expansion \eqref{eq:A_series} into Eq. \eqref{eq:E_rel}, we have 
\begin{align}\label{eq:E2_expansion}
  E^{(b)}&=
  \sum_{i<j}\expect{\left(1-\frac{\Delta\AO}{\expect{\AO}}+\left(\frac{\Delta\AO}{\expect{\AO}}\right)^2+\ldots\right)\frac{\left(\pOV_i-\pOV_j\right)^2}{2m\expect{\AO}}}+\expect{\VO}\,.
\end{align}
In this case, a naive power counting in $\expect{\AO}^{-1}$ breaks down because terms at a given order of the series are enhanced by factors of $\expect{\AO}$ as a direct consequence of applying Eq. \eqref{eq:A_promotion} to the expansion, i.e.,
\begin{align}
   \widetilde{E}^{(b)}_0 &= \expect{\VO}\,,\\
   \widetilde{E}^{(b)}_1 &= \widetilde{E}^{(b)}_0 + \sum_{i<j}\frac{\expect{\left(\pOV_i-\pOV_j\right)^2}}{2\expect{\AO}m}\notag\\
&=\widetilde{E}^{(b)}_0 + \expect{\TO}-\frac{\expect{\TO}}{\expect{\AO}}+\frac{\expect{\TO\Delta\AO}}{\expect{\AO}}-\sum_{i<j}\frac{\expect{\pOV_i\cdot\pOV_j}}{\expect{\AO}m}\label{eq:A_lo}\,,
\end{align}
and
\begin{align}\label{eq:A_nlo}
   \widetilde{E}^{(b)}_2&=\widetilde{E}^{(b)}_1-\sum_{i<j}\frac{\expect{\left(\pOV_i-\pOV_j\right)^2\Delta\AO}}{2\expect{\AO}^2m}\notag\\
  &=\widetilde{E}^{(b)}_1-\frac{\expect{\TO\Delta\AO}}{\expect{\AO}}+\frac{\expect{\TO\Delta\AO}}{\expect{\AO}^2}-\frac{\expect{\TO(\Delta\AO)^2}}{\expect{\AO}^2}\notag\\
  &\quad+\sum_{i<j}\frac{\expect{\pOV_i\cdot\pOV_j\Delta\AO}}{\expect{\AO}^2m}\,.
\end{align}
The term $\expect{\TO}$ that is formally of order $\expect{\AO}^{0}$ first appears in $\widetilde{E}^{(b)}_1$, a linear term in $\widetilde{E}^{(b)}_2$, and so on. Comparing $E^{(a)}_1$ with $\widetilde{E}^{(b)}_1$, we note that 
\begin{equation}\label{eq:tda}
  \widetilde{E}^{(b)}_1-E^{(a)}_1=\frac{\expect{\TO\Delta\AO}}{\expect{\AO}} \,,
\end{equation}
i.e., if we simply replace $\AO$ with the number $\expect{\AO}$ in Eqs. \eqref{eq:E_obtb} and \eqref{eq:E_rel}, the functionals are no longer equivalent!

To restore a proper power counting to the expansion of $E^{(b)}$, we first note that the enhanced linear term appearing in $\widetilde{E}^{(b)}_2$ exactly cancels the one in Eq. \eqref{eq:tda}. Likewise, an enhanced quadratic term in $\widetilde{E}^{(b)}_3$ cancels $\expect{\TO(\Delta\AO)^2}/\expect{\AO}^2$ in Eq. \eqref{eq:A_nlo}, and similar cancellations occur for all higher orders. The cancellation can be enforced explicitly if we define
\begin{equation}\label{eq:E_rel_correct}
   E^{(b)}_n = \widetilde{E}^{(b)}_n + (-1)^n\frac{\expect{\TO(\Delta\AO)^n}}{\expect{\AO}^n} \,,
\end{equation}
and applying Eq. \eqref{eq:A_promotion} we find that
\begin{equation}
   E^{(a)}_n = E^{(b)}_n\,, 
\end{equation}
i.e., the functionals $E^{(a)}$ and $E^{(b)}$ are identical at any given order of the expansion.

\subsection{Hartree-Fock-Bogoliubov Approximation}
We now apply the expansion developed in the previous subsection to the HFB approximation \cite{Ring:1980}, i.e., we assume that $\ket{\Psi}$ is a quasi-particle Slater determinant and introduce the density matrix
\begin{equation}\label{eq:def_rho}
  \rho_{kk'} = \dmatrixe{\Psi}{\ccO_{k'}\cO_k}
\end{equation}
and the pairing tensor
\begin{equation}\label{eq:def_kappa}
  \kappa_{kk'} = \dmatrixe{\Psi}{\cO_{k'}\cO_k}\,.
\end{equation}

We first consider the one- plus two-body form of the intrinsic kinetic energy. At next-to-leading order, the functional reads
\begin{align}
   E^{(a)}_1&=\left(1-\frac{1}{\expect{\AO}}\right)\sum_{kk'}\matrixe{k}{\tO}{k'}\rho_{k'k}\notag\\
     &\hphantom{=}+\frac{1}{2}\sum_{kk'qq'}\matrixe{kq}{\vO-\frac{\pOV_1\cdot\pOV_2}{\expect{\AO}m}}{k'q'}\rho_{q'q}\rho_{k'k}\notag\\
     &\hphantom{=}+\frac{1}{4}\sum_{kk'qq'}\matrixe{kk'}{\vO-\frac{\pOV_1\cdot\pOV_2}{\expect{\AO}m}}{qq'}\kappa^*_{kk'}\kappa_{qq'}\,,
\end{align}
where $\tO$ is the single-particle kinetic energy, and we assume that all two-body matrix elements are antisymmetrized in the following. Varying $E^{(a)}_1$ w.r.t. the density matrix, we obtain the particle-hole field
\begin{align}
   h^{(a)}_{kk'}
     &=\left(1\!-\!\frac{1}{\expect{\AO}}\right)\!\matrixe{k}{\tO}{k'} +\!
     \sum_{qq'}\matrixe{kq}{\vO-\frac{\pOV_1\cdot\pOV_2}{\expect{\AO}m}}{k'q'}\rho_{q'q}\notag\\
   &\quad+ \left(\frac{\expect{\TO}}{\expect{\AO}^2}+\sum_{i<j}\frac{\expect{\pOV_i\cdot\pOV_j}}{\expect{\AO}^2m}\right)\delta_{kk'}\,, 
\end{align}
where we have used 
\begin{equation}\label{eq:invA_var}
   \partd{}{\rho_{k'k}}\frac{1}{\expect{\AO}}=
     -\frac{1}{\expect{\AO}^2}\partd{}{\rho_{k'k}}\sum_j\rho_{jj}=-\frac{1}{\expect{\AO}^2}\delta_{kk'}\,.
\end{equation}
Varying the energy with respect to the pairing tensor, we find that the pairing field is given by
\begin{equation}
   \Delta^{(a)}_{kk'}=\frac{1}{2}\sum_{qq'}\matrixe{kk'}{\vO-\frac{\pOV_1\cdot\pOV_2}{\expect{\AO}m}}{qq'}\kappa_{qq'}\,.
\end{equation}

If we start from the pure two-body form \eqref{eq:tint_rel} of the intrinsic kinetic energy, we follow the analysis in the previous section and apply \eqref{eq:E_rel_correct} to obtain the NLO functional
\begin{equation}
   E^{(b)}_1 = \frac{2}{\expect{\AO}}\sum_{i<j}\frac{\expect{\qOV^2_{ij}}}{2\mu} - \frac{\expect{\TO\Delta\AO}}{\expect{\AO}} + \expect{\VO}\,.
\end{equation}
The expectation value of the correction term is
\begin{equation}
   \expect{\TO\Delta\AO}=2\sum_{kk'} \matrixe{k}{\tO}{k'}\left(\rho_{k'k}-\rho^2_{k'k}\right)\,,
\end{equation}
but due to the properties of the Bogoliubov transformation \cite{Ring:1980}
\begin{equation}
   \rho-\rho^2 = -\kappa\kappa^*\,,
\end{equation}
and there is an ambiguity regarding to which field the correction term contributes when $E^{(b)}$ is varied. To make contact with related works in the literature (see \cite{Hergert:2009nu} and Refs. therein) as well as the HF treatment of Ref. \cite{Khadkikar:1974}, we split it evenly between the particle-hole and particle-particle channels and express the latter contribution in a manifestly real form:
\begin{align}\label{eq:tda_choice}
  \expect{\TO\Delta\AO}&=\sum_{kk'} \matrixe{k}{\tO}{k'}\left(\rho_{k'k}-\rho^2_{k'k}\right) \notag\\
    &+ \frac{1}{2}\sum_{kk'}\left(\matrixe{k}{\tO}{k'}(\kappa\kappa^*)_{k'k} + \matrixe{k}{\tO}{k'}^*(\kappa^*\kappa)_{k'k}\right)\,.
\end{align}
We stress that the specific choice \eqref{eq:tda_choice} only affects \emph{unobservable quantities} like the particle-hole and pairing fields, while observables like the energy expectation value or the energy differences discussed in Sect. \ref{sec:results} do not depend on it at all.
 
Noting that
\begin{equation}
   \frac{2}{\expect{\AO}^2}\sum_{i<j}\frac{\expect{\qOV_{ij}^2}}{m}-\frac{\expect{\TO\Delta\AO}}{\expect{\AO}}=
\frac{\expect{\TO}}{\expect{\AO}}-\frac{\expect{\TO}}{\expect{\AO}^2}-\sum_{i<j}\frac{\expect{\pOV_i\cdot\pOV_j}}{\expect{\AO}^2m}
\end{equation}
(cf. Eq.\eqref{eq:A_lo}), we find that the particle-hole field is given by
\begin{align}\label{eq:h_rel}
   h^{(b)}_{kk'}&=\sum_{qq'}\matrixe{kq}{\frac{2\qOV_{12}^2}{\expect{\AO}m}+\vO}{k'q'}\rho_{q'q}-\frac{\expect{\TO}}{\expect{\AO}}\delta_{kk'}\notag\\
  &\hphantom{=}-\frac{1}{\expect{\AO}}\left(\!\matrixe{k}{\tO}{k'}-\!\sum_q\left(\matrixe{k}{\tO}{q}\rho_{qk'}+\rho_{kq}\matrixe{q}{\tO}{k'}\right)\!\right)\notag\\
   &\hphantom{=}- \left(\frac{\expect{\TO}}{\expect{\AO}}-\frac{\expect{\TO}}{\expect{\AO}^2}-\sum_{i<j}\frac{\expect{\pOV_i\cdot\pOV_j}}{\expect{\AO}^2m}\right)\delta_{kk'}\,, 
\end{align}
and for the pairing field, we obtain
\begin{align}\label{eq:d_rel}
  \Delta^{(b)}_{kk'}&=\frac{1}{2}\sum_{qq'}\matrixe{kk'}{\frac{2\qOV_{12}^2}{\expect{\AO}m}+\vO}{qq'}\kappa_{qq'}\notag\\
   &\hphantom{=}+\frac{1}{\expect{\AO}}\sum_{q}\left(\matrixe{k}{\tO}{q}\kappa_{qk'}+\kappa_{kq}\matrixe{q}{\tO}{k'}^*\right)\,.
\end{align}

Assuming that the single-particle states satisfy
\begin{equation}
   \braket{k}{k'}=\delta_{kk'}\,,
\end{equation}
we obtain the following relation for the antisymmetrized matrix element of the relative kinetic energy:
\begin{align}\label{eq:qme}
   \frac{2}{m}&\matrixe{kq}{\qOV^2_{12}}{k'q'}\notag\\
   &=\matrixe{k}{\tO}{k'}\delta_{qq'}-\matrixe{k}{\tO}{q'}\delta_{qk'}
    +\matrixe{q}{\tO}{q'}\delta_{kk'}\notag\\
   &\hphantom{=}-\matrixe{q}{\tO}{k'}\delta_{kq'}
    -\frac{1}{m}\matrixe{kq}{\pOV_1\cdot\pOV_2}{k'q'}\,.
\end{align}
Plugging this into Eqs. \eqref{eq:h_rel} and \eqref{eq:d_rel}, we find that
\begin{align}
   h^{(b)}_{kk'}&=h^{(a)}_{kk'}\,,\label{eq:h1h2}\\ 
  \Delta^{(b)}_{kk'}&=\Delta^{(a)}_{kk'}\,,
\end{align}
for the NLO functionals $E^{(a)}_1$ and $E^{(b)}_1$ and the specific choice \eqref{eq:tda_choice} for the correction term. 

The equivalence of the fields guarantees that a solution of the HFB equations for the functional $E^{(a)}_1$ will also solve the equations for $E^{(b)}_1$ and vice versa:
\begin{equation}\label{eq:HFB_eq}
   \comm{\mathcal{H}^{(a)}}{\mathcal{R}}=\comm{\mathcal{H}^{(b)}}{\mathcal{R}}=0\,.
\end{equation}
Here, $\mathcal{H}$ and $\mathcal{R}$ are the usual HFB Hamiltonian and generalized density matrices \cite{Ring:1980},
\begin{equation}
   \mathcal{H}=\begin{pmatrix} h - \lambda & \Delta \\ -\Delta^* & -h^* + \lambda \end{pmatrix}\,,\quad
   \mathcal{R}=\begin{pmatrix} \rho & \kappa \\ -\kappa^* & 1-\rho^* \end{pmatrix}\,,
\end{equation}
and the Lagrange multiplier satisfies
\begin{equation}
   \lambda = \lambda^{(a)} = \lambda^{(b)}\,.
\end{equation}
In this context, some additional remarks are in order. Since the expansion parameter $\expect{\AO}^{-1}$ is directly linked to the variational degrees of freedom $\rho_{kk'}$, the derivatives \eqref{eq:invA_var} generate $\expect{\AO}^{-2}$ terms in the fields that cause global shifts in the diagonal matrix elements of $h$, i.e., the underlying single-particle spectrum. In contrast to $\expect{\AO}^{-2}$ contributions that arise from varying the NNLO functionals $E^{(a/b)}_2$, these global shifts are \emph{state-independent}, and can be absorbed in the Lagrange multiplier $\lambda$ in a self-consistent calculation. If they are included implicitly in $\lambda$, one cannot directly interpret $\lambda$ as the Fermi energy of the system, as it is usually done in the literature (see e.g. \cite{Dobaczewski:1995bf}).

For higher orders of the expansion, the explicit evaluation of the expectation values $\expect{\TO\Delta\AO^n}$ and $\expect{\pOV_i\cdot\pOV_j\Delta\AO^n}$ occurring in higher-order functionals $E^{(a/b)}_n$ becomes increasingly cumbersome and time-consuming in practical calculations. Fortunately, the expansion of the energy functional converges rapidly, and it is sufficient to truncate the expansion of the energy functional at the linear order in practical calculations, as demonstrated in Sect. \ref{sec:results}.

\subsection{\label{sec:hf}Hartree-Fock Limit}
Starting from the analysis of the HFB approximation in the previous section, we can easily take the limit of vanishing pairing correlations to obtain the Hartree-Fock limit. Since the HF ground state $\ket{\Psi}$ is an eigenstate of $\AO$, we immediately see that
\begin{equation}
   \Delta\AO \ket{\Psi} = 0\,,
\end{equation}
and therefore all higher-order terms in the expanded functionals $E^{(1)}$ and $E^{(2)}$ automatically vanish. Of course, one could have directly used the eigenvalue equation \eqref{eq:A_eigen} and avoided the expansion in the first place. Moreover, this implies that the ``uncorrected'' HF functional 
\begin{equation}
  \widetilde{E}^{(b)}_{HF}
  = \frac{2}{A}\sum_{i<j}\frac{\expect{\qOV^2_{ij}}}{2\mu}  + \expect{\VO}
\end{equation}
defined in analogy to \eqref{eq:A_lo} yields the same energies as $E^{(a)}$ and $E^{(b)}$, while we obtain the relation
\begin{equation}\label{eq:h_KK}
   \widetilde{h}^{(b)}=h^{(a)}+\frac{1}{A}\left(\tO-(\tO\rho+\rho\tO)\right) + \frac{\expect{\TO}}{A}
\end{equation}
for the corresponding particle-hole Hamiltonian by moving the ``correction'' terms appearing in $h^{(b)}$ to the left-hand side of Eq. \eqref{eq:h1h2}. Equation \eqref{eq:h_KK} is exactly the relation Khadkikar and Kamble obtained by plugging \eqref{eq:qme} in the expression for $\widetilde{h}^{b}$ \cite{Khadkikar:1974}. From our analysis, we now see how this relation for the fields follows directly from the energy functional, and that the correction terms in particular are derivatives of \emph{vanishing energy contributions}. 

The HFB equations \eqref{eq:HFB_eq} for the functionals $E^{(a)}$ and $E^{(b)}$ are reduced to their HF counterparts, i.e., a HF solution $\rho$ obtained with either functional is also a valid solution for the other functional. In addition, it was demonstrated in \cite{Khadkikar:1974} that $\rho$ also solves the HF equations for the uncorrected functional $\widetilde{E}^{(b)}$, i.e.,
\begin{equation}
   \comm{h^{(a)}}{\rho}=\comm{h^{(b)}}{\rho}=\comm{\widetilde{h}^{(b)}}{\rho}=0\,.
\end{equation}

Finally, we point out one important caveat: the considerations of this subsection \emph{do not apply} if the equal-filling approximation (EFA) is employed to treat open-shell nuclei. While one can write down and solve HF-like equations for such a case, the EFA density matrix represents a statistical mixture rather than a genuine Slater determinant \cite{PerezMartin:2008yv}. In this case, we cannot use the eigenvalue equation \eqref{eq:A_eigen} to construct the energy functional, and we have to resort to the $1/\expect{\AO}$-expansion again, treating all expectation values in the statistical sense.

\subsection{\label{sec:vap}Particle-Number Projection}
Since nuclei are finite systems, one eventually wants to carry out a particle-number projection (PNP) to obtain a state with fixed particle number. The PNP energy functional is constructed from a quasi-particle Slater determinant that is explicitly projected onto the particle number $A$ via the hermitian projector $\PO_A$ 
\begin{equation}\label{eq:psi_proj}
   \ket{\Psi_A}=\PO_A\ket{\Psi}\,
\end{equation}
(see e.g. \cite{Ring:1980, Sheikh:1999yd} for details). Since $\ket{\Psi_A}$ satisfies the eigenvalue equation \eqref{eq:A_eigen}, an expansion is not required, just as in the HF case. All three previously defined functionals are equivalent when derived from \eqref{eq:psi_proj},
\begin{equation}\label{eq:energies}
   E^{(a)}_{PNP}=E^{(b)}_{PNP}=\widetilde{E}^{(b)}_{PNP}\,,
\end{equation}
but $\widetilde{E}^{(b)}_{PNP}$ will in general lead to different non-observable quantities like the projected fields $h_A$ and $\Delta_A$ or their individual contributions to the energy expectation value. In a variation after particle-number projection (VAP), the fields and densities obtained by solving the projected HFB equations \cite{Sheikh:1999yd} are associated with an auxiliary intrinsic state without physical meaning, whereas the expectation values of observables are well defined, even if they are $A$-dependent. Thus, only these expectation values of observables or derived quantities like separation energies (in the sense of energy differences) should be compared to experiment.

\section{\label{sec:results}Discussion \& Numerical Results}

\begin{figure}[t]
  \includegraphics[width=\columnwidth]{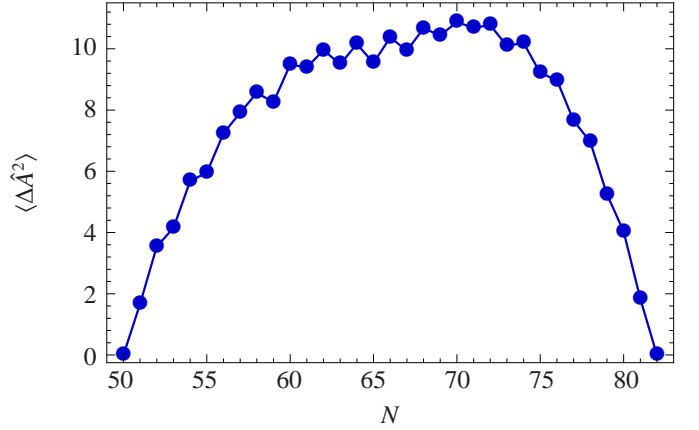}
  \caption{\label{fig:SnXXX_DA2}(Color online.) Particle number fluctuation in the tin isotopic chain.
  }
\end{figure}

To test the proposed expansion, we have performed spherical HFB calculations using the phenomenological Gogny D1S interaction \cite{Berger:1991}. We are employing a spherical harmonic oscillator configuration space, and explicitly minimize all energies with respect to the oscillator length $\aHO$; more details can be found in Ref. \cite{Hergert:2009nu}. Unless noted otherwise, we include 13 major oscillator shells in our calculations, which leads to a satisfactory convergence in the considered cases. Odd nuclei are treated in a self-consistent blocking method \cite{Ring:1980}, where the odd nucleon
is distributed evenly over all magnetic substates of a given $j$-shell according to the equal-filling approximation (see e.g. \cite{PerezMartin:2008yv}).

Let us first consider the convergence of the $1/\expect{\AO}$-expansion. Formally, an $A$-quasi-particle HFB state can contain states with sharp particle numbers from $0,\ldots,2A$. In the extreme cases, this would mean that the operator $\Delta\AO/\expect{\AO}$ in Eq. \eqref{eq:A_series} could acquire the operator norm $1$ on the space of HFB wavefunctions, causing the breakdown of the series expansion. In the nuclear many-body problem, this breakdown could only occur before the first major shell is fully occupied, and for these very light nuclei the use of mean-field methods is questionable in the first place. At the major shell closure itself, the HFB wavefunction collapses onto the HF solution, and there is no need for an expansion. As we progress along the nuclear chart, we find that particle-number fluctuations only occur within a given major shell, as shown exemplary for the tin isotopes with $N=50,\ldots,82$ in Fig. \ref{fig:SnXXX_DA2}. This implies that the operator norm of $\Delta\AO/\expect{\AO}$ remains below $1$ in practical applications, and guarantees the convergence of the series.

\begin{figure}[t]
  \centering
  \includegraphics[width=0.95\columnwidth]{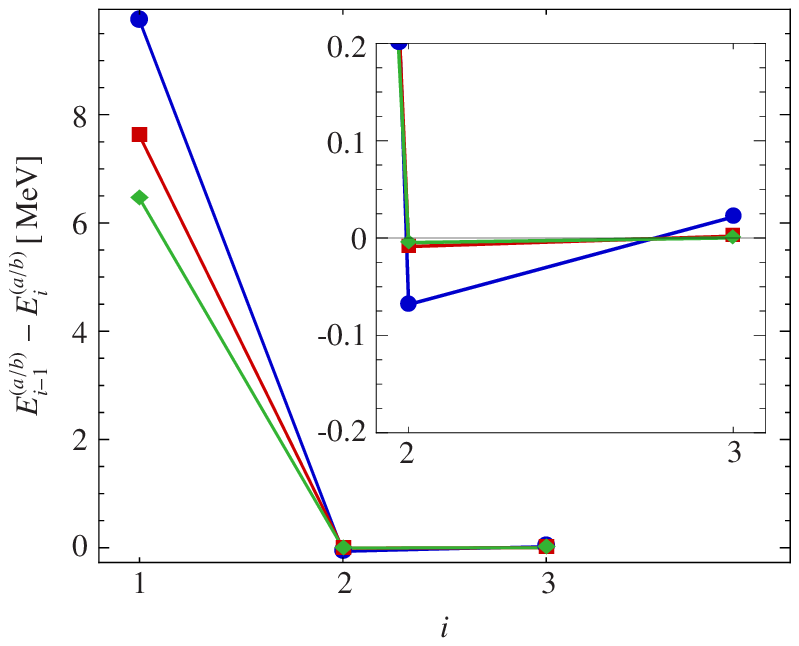}
  \caption{\label{fig:exp_conv}(Color online.) Convergence of the $1/\expect{\AO}$ expansion: ground-state energy difference $E^{(a/b)}_{i-1}-E^{(a/b)}_i$ as a function of the order $i$ for $\nuc{O}{18}$ \symbolcircle[FGBlue],
    $\nuc{Ni}{64}$\symbolbox[FGRed], and $\nuc{Sn}{120}$\symboldiamond[FGGreen].
  }
\end{figure}

To test the rate of the convergence, we display the quantity $E^{(a/b)}_{i-1}-E^{(a/b)}_i$ for the open-shell nuclei $\nuc{O}{18}$, $\nuc{Ni}{64}$, and $\nuc{Sn}{120}$ in Fig. \ref{fig:exp_conv}. We have picked these specific nuclei for their large values of $\expect{\Delta\AO^2}$ in the respective isotopic chains. We find that is essentially sufficient to include the linear terms in $1/\expect{\AO}$ in the energy functional. Beyond the linear order, the largest variations occur for $\nuc{O}{18}$, where the successive inclusion of terms up to third order causes changes of $100-200\,\keV$ (see the inset of Fig. \ref{fig:exp_conv}). As expected, the effect of higher orders rapidly diminishes with increasing masses, and amounts to a few $\keV$ for the nickel and tin isotopes. For this reason, we will consider all functionals at linear (i.e. next-to-leading) order, and drop the subscripts in the following.

\begin{figure}[t]
  \includegraphics[width=\columnwidth]{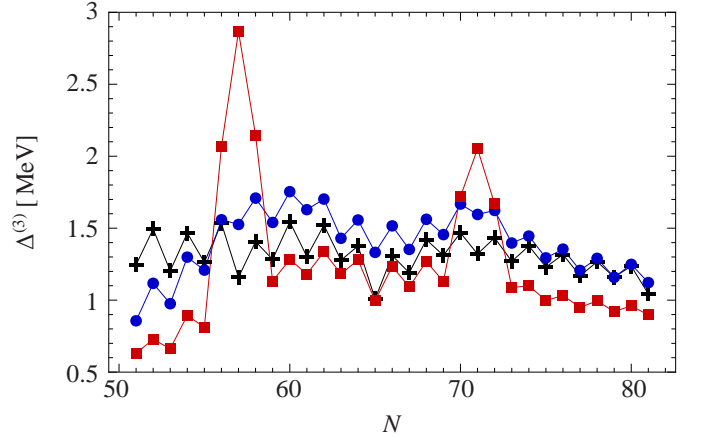}
  \caption{\label{fig:SnXXX_D3}(Color online.) Theoretical three-point binding energy differences in the tin isotopic chain from $E^{(a/b)}$\,\symbolcircle[FGBlue] and $\widetilde{E}^{(b)}$\,\symbolbox[FGRed]\,, compared to experimental values \symbolcross \cite{Audi:2002rp}.
  }
\end{figure}

Information about the effect of the intrinsic kinetic energy on the nuclear pairing correlations can be extracted by considering the three-point binding energy difference formula
\begin{equation}\label{eq:def_D3}
  \Delta^{(3)}(N)=(-1)^N\frac{1}{2}\left(E(N+1)-2 E(N) + E(N-1)\right)
\end{equation}
along an isotopic chain (or equivalently, an isotonic chain with $N$ replaced by $Z$). By calculating the ground states of odd nuclei self-consistently, we avoid the complications arising in perturbative analyses, as discussed in detail in \cite{Duguet:2001gr, Duguet:2001gs}. This is particularly relevant for $A$-dependent Hamiltonians \cite{Hergert:2009nu}, which add another layer of complexity to the perturbative approach. 

In Fig. \ref{fig:SnXXX_D3}, we compare the theoretical $\Delta^{(3)}$ for the properly constructed functional to the uncorrected functional $\widetilde{E}^{(b)}$ \eqref{eq:A_lo}. The latter are typically lower than the values for the proper functional by $200-400\,\keV$, with the exceptions occurring around the $1d_{5/2}$ and $1d_{3/2}$ sub-shell closures at $N=56$ and $N=70$, respectively. The gaps between the relevant sub-shell and the next available one are notably larger for $\widetilde{E}^{(b)}$ than for $E^{(a/b)}$, leading to more pronounced effects when the odd nucleon is added 

\begin{figure}[t]
  \includegraphics[width=\columnwidth]{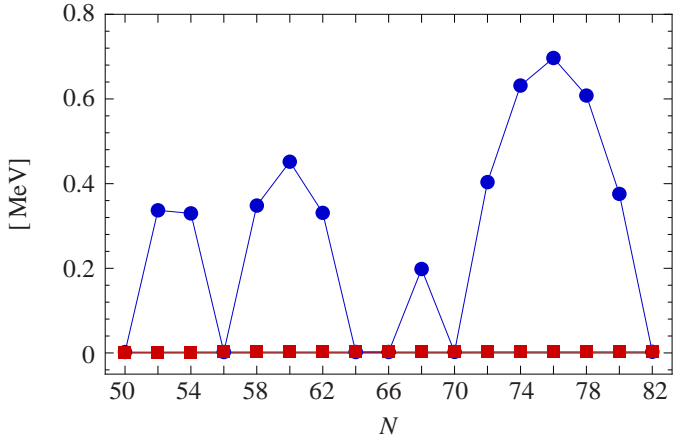}
  \caption{\label{fig:SnXXX_DEhf}(Color online.) Ground-state energy differences for the tin isotopes in HF+EFA: $\widetilde{E}^{(b)}_{HF}-E^{(a)}_{HF}$
  \symbolcircle[FGBlue] and $E^{(b)}_{HF}-E^{(a)}_{HF}$ \symbolbox[FGRed]. Calculations were done with 11 major shells.
  }
\end{figure}

In Sect. \ref{sec:hf}, we have pointed out that a spherical HF treatment of open-shell nuclei will suffer from the same problems as the HFB method due to the use of the equal-filling approximation. This is explicitly demonstrated for the tin isotopic chain in Fig. \ref{fig:SnXXX_DEhf}, where we have used the density matrix $\rho$ obtained by minimizing the functional $E^{(a)}_{HF}$ to calculate the energies $\widetilde{E}^{(b)}_{HF}$ and $E^{(b)}_{HF}$. At sub-shell closures, we have a genuine HF problem, and in this case \emph{all three} functionals are equivalent and $\rho$ is a solution for each set of HF equations \cite{Khadkikar:1974}. For open-shell nuclei, in contrast, the binding energies obtained with $\widetilde{E}^{(b)}_{HF}$ are reduced by several hundred $\keV$, and the difference is given by
\begin{equation}
   \frac{\expect{T\Delta\AO}}{\expect{\AO}}=\frac{1}{\expect{\AO}}\tr\left(\rho-\rho^2\right)
\end{equation}
to numerical accuracy.

\begin{figure}
  \centering
  \includegraphics[width=0.95\columnwidth]{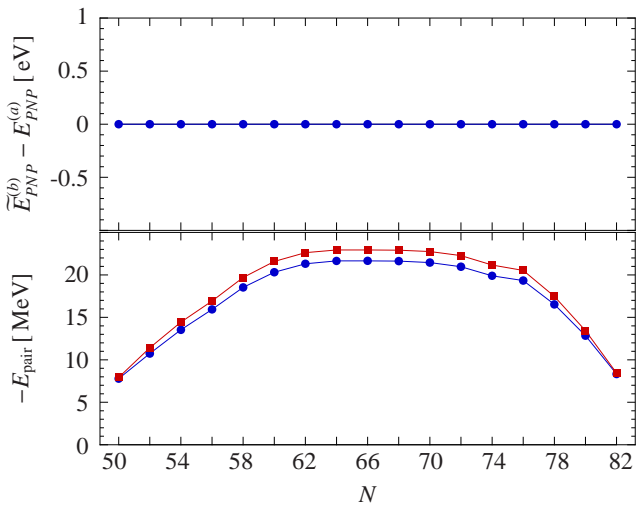}
  \caption{\label{fig:SnXXX_VAP}(Color online.) Top: Ground-state energy differences $\widetilde{E}^{(b)}_{PNP}-E^{(a)}_{PNP}$ of the tin isotopes from VAP calculations (note the scale in $\eV$). Bottom: Total pairing energies of the tin isotopes from VAP calculations with $E^{(a)}_{PNP}$ \symbolcircle[FGBlue] and $\widetilde{E}^{(b)}_{PNP}$ \symbolbox[FGRed]. All calculations were done with 11 major shells.
 }
\end{figure}

Finally, we compare the tin ground-state energies obtained with $\widetilde{E}^{(b)}_{PNP}$ to those of $E^{(a/b)}_{PNP}$ in calculations where the variation is carried out after particle-number projection (VAP). We are aware that the VAP method is highly problematic for density-dependent interactions like Gogny D1S (see e.g. \cite{Duguet:2008rr} and references therein). Since we do not aim for a comparison with experiment in the present discussion, we assume that the potential energies of both calculations will be equally affected by any spurious behavior due to the density-dependence, hence any differences in the VAP energy expectation values are due to the different forms of the kinetic energy. Looking at the top half of Fig. \ref{fig:SnXXX_VAP} and noting the $\eV$ scale, we find no such differences: the tin ground-state energies obtained in VAP calculations with either the properly constructed functional $E^{(a/b)}_{PNP}$ or the naive, uncorrected functional $\widetilde{E}^{(b)}_{PNP}$ are identical,  confirming the discussion of Sect. \ref{sec:vap}. At the same time, the comparison of the pairing energies in the bottom half of Fig. \ref{fig:SnXXX_VAP} illustrates that the \emph{individual} particle-hole and particle-particle contributions to the total energy expectation value may well be different in both calculations.

\section{Conclusions}
We have presented a detailed analysis of the $\AO$-dependent intrinsic kinetic energy operator in theories with and without particle-number conservation, in particular the mean-field type Hartree-Fock-Bogoliubov method and its number-conserving extension via particle-number projection. We have shown that a naive treatment where the number operator $\AO$ is replaced by its expectation value $\expect{\AO}$ causes discrepancies in expectation values obtained with the otherwise equivalent operator forms of $\Tint$. We have developed a systematic expansion to fix this problem, but we emphasize that this expansion does \emph{not} restore, nor is it intended to restore, either the particle number or translational symmetries of the nucleus.

Our discussion provides an \emph{a posteriori} justification for using the one- plus two-body form of the intrinsic kinetic energy since it is automatically consistent with the power counting of the developed expansion. As a byproduct, we also clarify how differences and ambiguities in non-observable quantities which had been discussed in the context of the HF approximation \cite{Khadkikar:1974} arise systematically in the presented framework. While we have discussed the specific case of the intrinsic kinetic energy operator in the present article, we point out that the same treatment should be applied to all $\AO$-dependent observables. 

\section*{Acknowledgments}
We thank T. Duguet, N. Schunck, and T. Lesinski for useful discussions. This work is supported by the Deutsche Forschungsgemeinschaft through contract SFB 634 and by the Helmholtz International Center for FAIR within the framework of the LOEWE program launched by the State of Hesse. We thank the Institute for Nuclear Theory at the University of Washington for its hospitality and the Department of Energy for partial support during the initiation of this work.


\end{document}